\documentclass[11pt]{amsart}
\usepackage{geometry}                
\usepackage{graphicx}
\usepackage{amssymb}
\usepackage{epstopdf}
\usepackage{hyperref}
\pdfoutput=1
\title{Preferential Condensation of water droplets using hybrid hydrophobic-hydrophilic surfaces}
\author{Adam Paxson, Kripa K. Varanasi \\
Massachusetts Institute of Technology, Cambridge, MA\\
Tao Deng, Ming Hsu, Nitin Bhate\\
GE Global Research, Niskayuna, NY}

\begin{document}
\pagestyle{empty}
\maketitle
\begin{abstract}
We present a hybrid microtextured surface with heterogeneous hydrophilic-hydrophobic regions for condensing water vapor while maintaining anti-wetting behavior. Fluid dynamics videos are shown demonstrating the difference between condensation on a homogeneous hydrophobic structure and the same structure with hybrid wetting regions. 
\end{abstract}
\section{Brief Explanation of Video Submission}
It is well-known that superhydrophobic surfaces are able to resist wetting, either by sessile drops or by drops impacting with an appreciable velocity. This anti-wetting behavior is possible due to the roughness features of the surface, which are able to support drops in a superhydrophobic Cassie state. In this state, the drops rest on the tops of the roughness features, with a vapor layer remaining between the drop and the surface. 

This superhydrophobic Cassie state is difficult to maintain during condensation. Water droplets will commonly nucleate and begin growing within the roughness features. As the condensation process proceeds, small, nucleated droplets grow and coalesce into neighboring drops so that when the drop has grown to macroscopic size, it is in a highly-pinned Wenzel state. These drops are difficult to remove, and thus the anti-wetting properties of the surface are lost. The video shows a surface patterned with 10 $\mu$m square silicon microposts with a height of 10 $\mu$m, coated in fluorosilane with an intrinsic contact angle of $120\,^{\circ}$. The surface is exposed to water vapor at $2.4\,^{\circ}\mathrm{C}$ and 800 Pa.

If a surface is patterned with distinct hydrophilic and hydrophobic areas, the condensation process proceeds differently. In this case, nucleation and growth occurs preferentially on the hydrophilic areas, while the hydrophobic areas experience little to no nucleation or growth of droplets. On a smooth surface, this effect is quite evident in the rapid growth of droplets onto hydrophilic areas as compared to hydrophobic areas. When applied to a structure with a micropillar texture, the tops of the pillars can be rendered hydrophilic so that they experience preferential condensation. Nucleation occurs at the tops of the posts more readily than within the posts, so that drops do not grow within the roughness features. Instead, the condensed drops remain only at the tops of the posts, leading to a superhydrophobic surface that retains anti-wetting behavior even under condensation. 

\section{Video Submissions}
\href{http://dl.dropbox.com/u/5995887/PrefCondensation_presentation.mp4}{Video 1} is a presentation-quality version of the fluid dynamics video. 

\href{http://dl.dropbox.com/u/5995887/PrefCondensation_web.mp4}{Video 2} is a web-quality version of the fluid dynamics video. 

\section{References}
K. K. Varanasi, M. Hsu, N. Bhate, W. Yang, T. Deng, ÒSpatial Control in the Heterogeneous Nucleation of Water,Ó Applied Physics Letters, accepted for publication July, 2009.

T. Deng, K. K. Varanasi, M. Hsu, N. Bhate, C. Keimel, J. Stein, M. Blohm ÒNonwetting of Impinging Droplets on Textured SurfacesÓ Applied Physics Letters, 94, 133109, 2009.
\end{document}